\documentclass[runningheads]{llncs}
\usepackage{graphicx}
\usepackage{graphics} 
\usepackage{amsmath} % assumes amsmath package installed
\usepackage{amssymb}  % assumes amsmath package installed
\usepackage{tabularx}
\usepackage{bm,upgreek}
\usepackage{amsfonts}
\usepackage{caption}
\usepackage{fontenc}
\usepackage{subcaption}
\usepackage{float}
\usepackage[hidelinks]{hyperref}
\graphicspath{{figures/}}
\usepackage{cite}
\usepackage{xcolor}

\begin{document}
\title{Stigmergy-based collision-avoidance algorithm for self-organising swarms}
\author{Paolo Grasso\inst{1,2}\orcidID{0000-0003-1195-1859} \and \\
Mauro Sebasti\'an Innocente\inst{1,3}\orcidID{0000-0001-8836-2839}}
\authorrunning{P. Grasso and M.S. Innocente}
\institute{Autonomous Vehicles \& Artificial Intelligence Laboratory (AVAILab),\\
	Centre for Future Transport and Cities, Coventry University, Coventry, UK\\
	\url{https://www.availab.org/} \and
	\email{grassop@coventry.ac.uk} \and
	\email{Mauro.S.Innocente@coventry.ac.uk}}

\maketitle{5$^\text{th}$ Int. Conference on Computational Vision and Bio-Inspired Computing}  % typeset the header of the contribution
\begin{abstract}
Real-time multi-agent collision-avoidance algorithms comprise a key enabling technology for the practical use of self-organising swarms of drones. This paper proposes a decentralised reciprocal collision-avoidance algorithm, which is based on stigmergy and scalable. The algorithm is computationally inexpensive, based on the gradient of the locally measured dynamic cumulative signal strength field which results from the signals emitted by the swarm. The signal strength acts as a repulsor on each drone, which then tends to steer away from the \textit{noisiest} regions (cluttered environment), thus avoiding collisions. The magnitudes of these repulsive forces can be tuned to control the relative importance assigned to collision avoidance with respect to the other phenomena affecting the agent's dynamics. We carried out numerical experiments on a self-organising swarm of drones aimed at fighting wildfires autonomously. As expected, it has been found that the collision rate can be reduced either by decreasing the cruise speed of the agents and/or by increasing the sampling frequency of the global signal strength field. A convenient by-product of the proposed collision-avoidance algorithm is that it helps maintain diversity in the swarm, thus enhancing exploration.

\keywords{Decentralised \and Multi-agent \and Autonomous \and Wildfires.}
\end{abstract}
\section{Introduction}

Multi-agent systems such as drone swarms consist of a multitude of decision-making individuals that interact directly and indirectly within the environment in order to achieve one or more predefined goals. In such systems, one of the most critical issues to be dealt with is collision avoidance. In this paper, we adhere to the classification of Collision-Avoidance Systems (CASs) suggested in \cite{Albaker2010} and \cite{Yasin2020}, which identifies the following five classes:
\begin{enumerate}
	\item Predefined collision avoidance.
	\item Protocol-based collision avoidance.
	\item Optimised escape trajectory approaches.
	\item E-field / potential field methods.
	\item Miscellaneous.
\end{enumerate}

The first two categories encompass the original and most trivial approaches. Predefined collision avoidance follows a fixed set of unchangeable rules requiring no computation, whilst protocol-based collision avoidance relies on both a set of rules and continuous exchange of information among agents with regards to their states. The former results in a faster reaction time whereas the latter is safer. The most advanced CASs fall within the remaining three categories.

The optimised escape trajectory approach consists of solving an optimisation problem that combines the drone kinematic model with a set of physical and behavioural constraints. The following decentralised multi-agent algorithms fall within this group: 1)~the Reciprocal Velocity Obstacles (RVO) and the derived Optimal Reciprocal Collision-Avoidance (ORCA) algorithms \cite{Berg2008}; 2)~the Cooperative Dynamic (CoDy) algorithm \cite{Regele2006}, which is able to solve dead-lock situations; 3)~the Context-Aware Route Planning (CARP) algorithm \cite{Blesing2017,Mors2011}, which is a graph routing algorithm aimed at finding the shortest trajectory and avoiding collisions; and 4)~other optimisation-based algorithms such as \cite{Osten2014} and \cite{Zhang2017}.

Among the E-field / potential field methods, the original ones tend to get drones randomly trapped in local minima (e.g. \cite{Khatiba}), though this issue has now been dealt with (e.g. \cite{Kim1992,Chang}). The advantages are their low computational cost and short reaction time, whilst two major disadvantages still persist: 1)~the requirement of high-level flight guidance, and 2)~the potential occurence of hard-to-perform sharp discontinuities in the commanded manoeuvre.

The miscellaneous category includes approaches such as a \emph{sense-and-avoid} algorithm for structured multi-agent systems, in which a leader is followed blindly by other agents \cite{Soriano2013,Yasin2020}. Other researchers explore more exotic solutions such as a two-stage reinforcement learning approach for multi-drone systems under imperfect sensing \cite{Wang2020}. Their aim is to train a policy to plan a collision-free trajectory by leveraging local noisy observations. Others try to infer the state of the overall swarm through stigmergy, assuming this would be useful for collision avoidance purposes. For example, an approach based on anticipatory stigmergic collision avoidance (ASCA) under noise is proposed in \cite{Osten2014}, which consists of using pheromone information in a rather unusual fashion: instead of leaving a trail of pheromones over past positions, all drones in the swarm share information in the same indirect manner but about future intended positions instead. The drones will then optimise their trajectories in order to avoid locations with high concentrations of pheromone.

Numerous CASs have been proposed in the literature, the majority of which are suitable for Autonomous Ground Vehicles (AGVs) only. This is because CASs typically rely on two main assumptions \cite{Osten2014}: 1)~\emph{stationary vehicles} (i.e. vehicles can stop suddenly and remain still indefinitely) and 2)~\emph{perfect information} (accurate noiseless perception). This is certainly not the case for Unmanned Aerial Vehicles (UAVs) or Unmanned Underwater Vehicles (UUVs), which might need to move with limited awareness of other vehicles' locations due to vision obstructions (e.g. smoke, cluttered environments, cloudy water) as well as high variance and bias of the positioning system being used.

In this paper, we propose a reactive decentralised stigmergy-based reciprocal collision-avoidance algorithm to be implemented on a self-organising swarm of drones. The algorithm is tested on a two-dimensional autonomous firefighting system based on \cite{InnGra18,Innocente2019} aimed at suppressing a simulated wildfire \cite{Grasso2020}. Formulation allows for a relatively seamless generalisation to three dimensions, which is beyond the scope of this paper. The proposed multi-agent collision-avoidance algorithm was inspired by the E-field / potential field methods \cite{Albaker2010}, the charged particles method \cite{BlaBen02}, and the classical Reynold's rules \cite{Rey87,BraETAL18}. However, these methods are based on direct communication and inter-agent distance computations. Conversely, the proposed algorithm is indirect via stigmergy using information that is continuously available and dynamically changing in the environment: the cumulative strength of a continuously broadcasted signal emitted by every drone. Thus, the drone behavioural rules are modified by adding a repulsor that pushes the trajectory in the direction opposite to the signal gradient. Numerical experiments are performed in order to investigate the effects of \emph{cruise speed} and \emph{sampling frequency} on the performance of the algorithm. Finally, future work towards generalising the algorithm is suggested.

\section{Multi-Agent Collision Avoidance Based on Stigmergy}
A stigmergy-based multi-agent collision-avoidance algorithm is developed as an efficient and reliable alternative to multi-agent algorithms in the literature which require inter-agent direct communications. This approach is expected to be more reliable in real-world applications such as firefighting, since the required information is measured from --and in turn modifies-- the environment therefore avoiding the loss of direct agent-to-agent communications. Given the early stages of this research, comparisons against other methods are not carried out in this paper. Instead, the aim is to demonstrate that the proposed method works and to investigate how \emph{cruise speed} and \emph{signal sampling time} may affect its performance.

Considering that the swarm-dynamics model used \cite{Innocente2019} is based on the Particle Swarm Optimisation method, the proposed collision-avoidance algorithm is embedded within the particle (i.e. agent or drone) trajectory difference equation in the form of a repulsor. Since each agent has no information regarding the location of the others, indirect communication (i.e. stigmergy) comprises a highly desirable feature: communication among agents is carried out by measuring and modifying the surrounding signal strength field in the environment. Thus, the process consists of two main stages: 1)~sufficiently high frequency simultaneous sampling of the signal intensity in a few points surrounding the agent to calculate the gradient, and 2)~correction of the attractor by a convex combination of the collision-avoidance repulsor and the original attractor.

The original formulation of the target position of every drone is regularly estimated at a predefined frequency as follows \cite{Innocente2019}:
\begin{equation}
	\begin{cases}
		\mathbf{xt}^{(t+1)}_{i} = \mathbf{x}^{(t)}_{i} + \omega \cdot \left ( \mathbf{x}^{(t)}_{i} - \mathbf{x}^{(t-1)}_{i} \right ) + \bm{\upphi}^{(t)}_{i} \odot \left ( \mathbf{p}^{(t)}_{i} - \mathbf{x}^{(t)}_{i} \right )
		\\
		\bm{\upphi}^{(t)}_{i} = \phi_{\text{min}} + \mathbf{U_{(0,1)}} \cdot \left ( \phi_{\text{max}} -  \phi_{\text{min}} \right )\\
		\phi_{\text{max}}, \phi_{\text{min}} = f(\omega)
	\end{cases} 
	\label{eq-PSO}
\end{equation}
where $\mathbf{xt}^{(t)}_{i}$ and $\mathbf{x}^{(t)}_{i}$ are the target and actual positions of the $i^\text{th}$ drone, respectively, at time-step $t$; $\bm{\upphi}^{(t)}_{i}$ is the acceleration coefficient; $\mathbf{p}^{(t)}_{i}$ is the attractor; $\omega$ is the inertia weight; $\mathbf{U_{(0,1)}} = \left[U_{(0,1)},U_{(0,1)}\right]^\text{T}$ with $U_{(0,1)}$ being a random variable realised from a uniform distribution within (0,1); and $\odot$ represents the component-wise product. The bounds $\phi_{\text{min}} = (\omega+1)$ and $\phi_{\text{max}}=(\sqrt{\omega}-1)^2$ are defined in \cite{Inn21} and \cite{Innocente2019} to ensure oscillatory behaviour of the drones.

Note that $\mathbf{x}^{(t+1)}_{i}$ is generally not equal to $\mathbf{xt}^{(t+1)}_{i}$ due to constraints such as the maximum cruise speed. It is also important to note that $\bm{\upphi}^{(t)}_{i}$ takes different values for different drones, for different components, and for different time-steps because $U_{(0,1)}$ is resampled anew every time it is referenced. However, $U_{(0,1)}$ is replaced by a constant of 0.5 in this paper to obtain deterministic behaviour thus facilitating the study of other relevant parameters. If in recharging or water replenishment modes, the attractor is given by the location of the respective sources. In this paper, the attractor is modified by adding a repulsor associated to the signal strength field for collision avoidance purposes ($\mathbf{q}^{(t)}_{i}$ in \eqref{eq-attractormod}).

\begin{equation}
	\begin{cases}
		\mathbf{p}^{(t)\star}_{i} = \left ( 1 - k_{\text{ca}}\right ) \cdot \mathbf{p}^{(t)}_{i} +k_{\text{ca}} \cdot \mathbf{q}^{(t)}_{i} \\
		0 \leq k_{\text{ca}} < 1
	\end{cases} 
	\label{eq-attractormod}
\end{equation}

The repulsor $\mathbf{q}^{(t)}_{i}$ is function of the gradient of the signal strength field ($\sigma$):

\begin{equation}
	\begin{cases}
		{q}^{(t)}_{ij} =  x_{ij}^{(t)} -  k_{\sigma} \cdot \frac{\partial \sigma(\mathbf{x}^{(t)}_{i})}{\partial x_j} \quad j=1,2 \\
		\sigma(\mathbf{x}^{(t)}_{i}) = \Sigma^N_{m=1} \sigma_m(\mathbf{x}^{(t)}_{i})
	\end{cases}
	\label{eq-repulsor}
\end{equation}

The individual signal intensity fields ($\sigma_m$) are normalised and defined as a power law based on the distance from the $m^\text{th}$ source as in \eqref{eq-sigma}.

\begin{equation}
	\begin{matrix}
		\sigma_m(\mathbf{x}^{(t)}_{i}) = \begin{cases}
			r_{\text{ref}}^2 \cdot r_{im}^{-2} &  \quad r_{im} > r_{\text{ref}} \\
			1 & 0 \leq  r_{im} \leq r_{\text{ref}}
		\end{cases}\\
		r_{im} = \left \| \mathbf{x}^{(t)}_{i} - \mathbf{x}^{(t)}_{m} \right \|
	\end{matrix}
	\label{eq-sigma}
\end{equation}
where $r_{\text{ref}}$ is a reference radius of a circle delimiting the area of generation of the signal within each drone. Setting $r_{\text{ref}}$ approximately equal to the size of the drone or smaller ensures that the second condition in \eqref{eq-sigma} never occurs. The proportionality constants in \eqref{eq-attractormod} and \eqref{eq-repulsor} depend on the magnitude of the other attractors in \eqref{eq-PSO}, and therefore need to be tuned for the problem in hand. In this paper, $k_{\text{ca}} = 0.7$ and $k_{\sigma}=1000$. The gradient was calculated on a stencil of eight points surrounding the position of the drone on a structured grid with the same increment in both directions ($\Delta x = \Delta y$) as shown in \eqref{eq-stencil}.

\begin{equation}
	\begin{cases}
		\displaystyle \frac{\partial \sigma}{\partial x_j} \simeq \frac{d \sigma}{d x_i} \\ \\
		\displaystyle\frac{d \sigma}{d x_1} = \frac{1}{6 \Delta x} \cdot \left ( \sigma_{\text{E}} - \sigma_{\text{W}} + 0.5 \cdot (\sigma_{\text{NE}} + \sigma_{\text{SE}} - \sigma_{\text{SW}} - \sigma_{\text{NW}}) \right )\\ \\
		\displaystyle\frac{d \sigma}{d x_2} = \frac{1}{6 \Delta y} \cdot \left ( \sigma_{\text{N}} - \sigma_{\text{S}} + 0.5 \cdot (\sigma_{\text{NE}} + \sigma_{\text{NW}} - \sigma_{\text{SW}} - \sigma_{\text{SE}}) \right )
	\end{cases}
	\label{eq-stencil}
\end{equation}

\section{Numerical Experiments}

\begin{figure*}[ht!]
	\begin{subfigure}[b]{0.49\linewidth}\centering
		\includegraphics[width=0.97\linewidth]{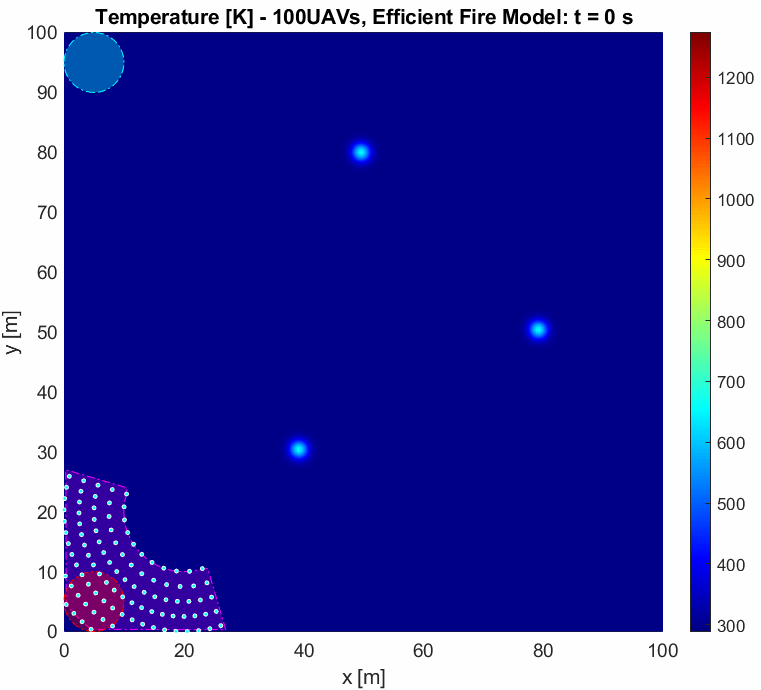} \caption{Initial condition\vspace{2mm}}
	\end{subfigure}
	\begin{subfigure}[b]{0.49\linewidth}\centering
		\includegraphics[width=0.97\linewidth]{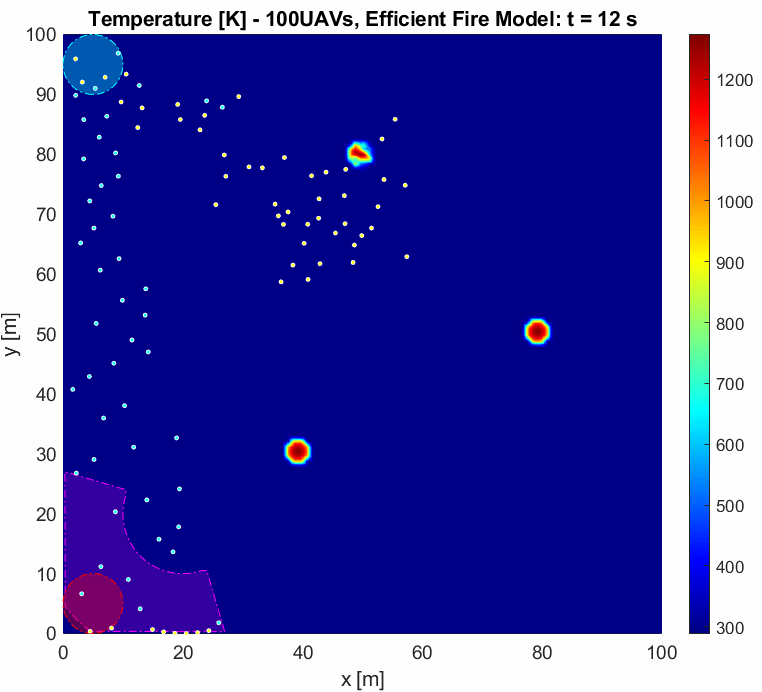} \caption{Water collection (ciano)\vspace{2mm}}
	\end{subfigure}
	
	\begin{subfigure}[b]{0.49\linewidth}\centering
		\includegraphics[width=0.97\linewidth]{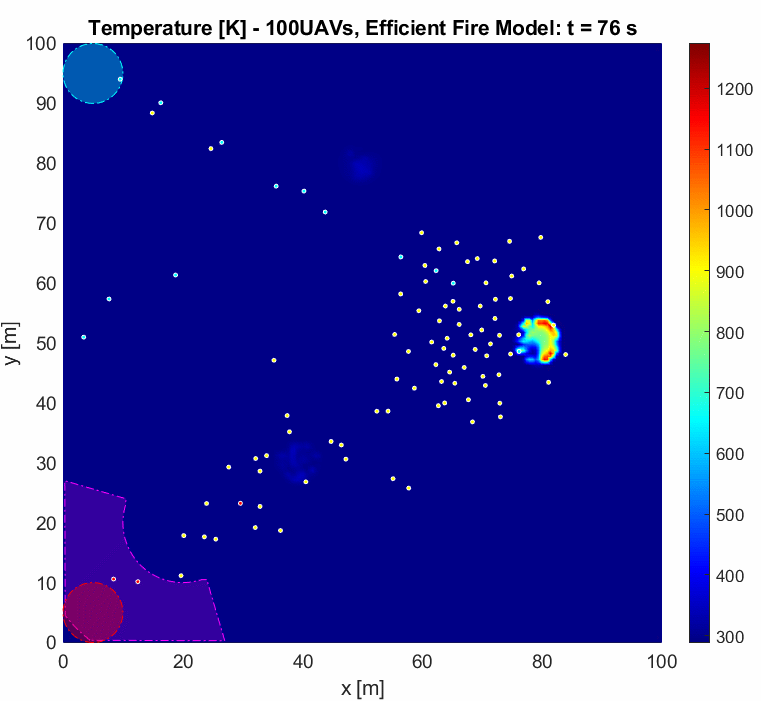} \caption{Recharging (red)\vspace{2mm}}
	\end{subfigure}
	\begin{subfigure}[b]{0.49\linewidth}\centering
		\includegraphics[width=0.97\linewidth]{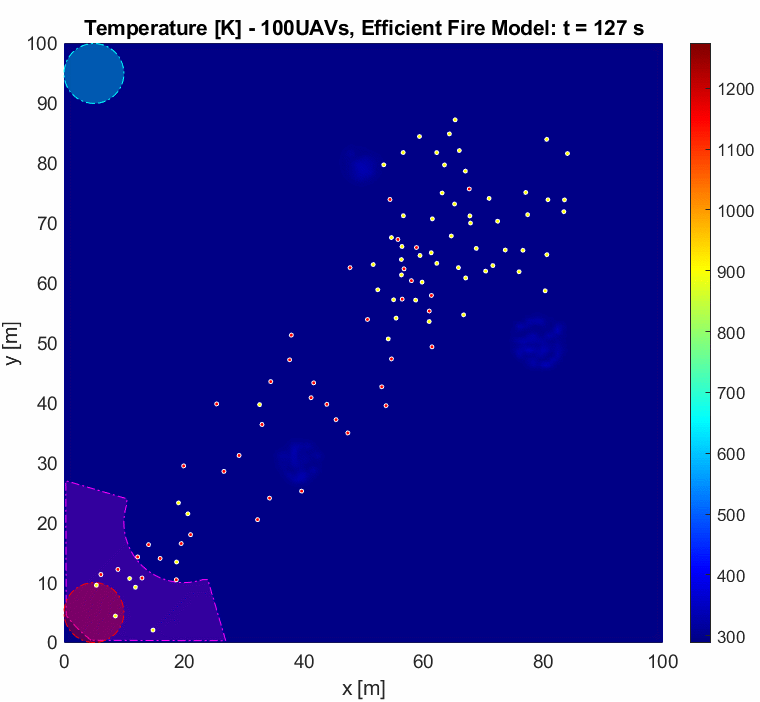} \caption{Fire extinction\vspace{2mm}}
	\end{subfigure}
	
	\begin{subfigure}[b]{0.49\linewidth}\centering
		\includegraphics[width=0.97\linewidth]{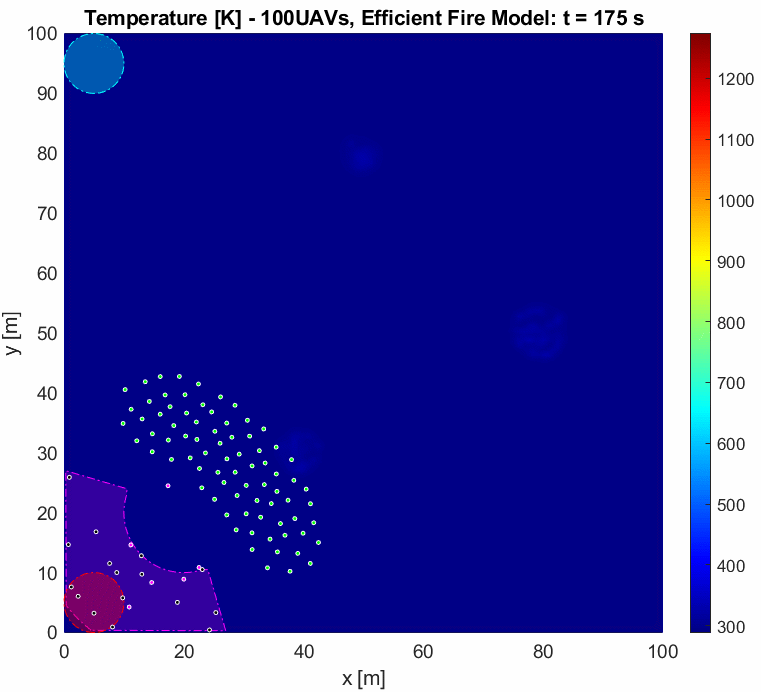} \caption{\centering Hovering (green) \& landing (magenta)\vspace{2mm}}
	\end{subfigure}
	\begin{subfigure}[b]{0.49\linewidth}\centering
		\includegraphics[width=0.97\linewidth]{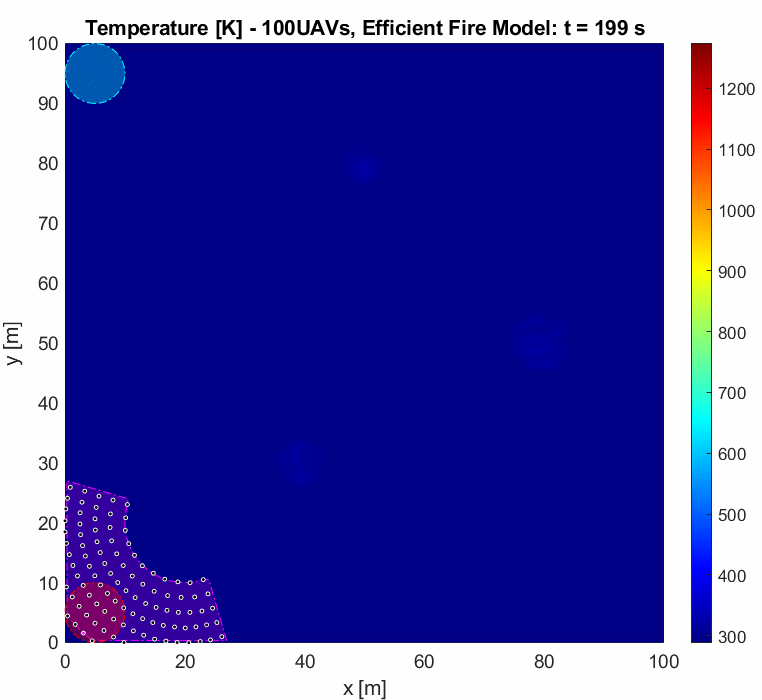} \caption{\centering Landed (black) on docking station\vspace{2mm}} 
	\end{subfigure}
	
	\caption{Top views of a 100m$\times$100m homogeneous distribution of combustible vegetation ignited in three locations (Fig.~a) and some representative instances for the main mission phases of a swarm of 100 firefighting drones. The colourmap represents the temperature field in Kelvin, ranging from ambient temperature (dark blue) to high flaming temperature (dark red). The red circle on the bottom-left of each figure shows the recharging area, the blue circle on the top-left corner shows the water source, whilst the \emph{arena polygon} shows the landing area.}
	\label{fig-phases}
\end{figure*}

The stigmergy-based multi-agent collision-avoidance algorithm developed in this paper is tested on a swarm of deterministically self-organising drones \cite{Innocente2019} fighting a simulated wildfire \cite{Grasso2020}. The collision-avoidance algorithm is always active while drones are airborne during any of the following mission phases for the simulated firefighting operations \cite{Innocente2019} in Figure \ref{fig-phases}:

\begin{itemize}
	\item \emph{Initialisation} \: Drones are positioned in their docking station with no water in their hull, the combustible vegetation fuel is distributed in the domain, and the wildfire is started via multiple ignition points (e.g. three in Fig.~\ref{fig-phases}).
	\item \emph{Firefighting} \: Drones search for fires by recording their best experience and sharing it with the entire swarm. They will pour a predefined amount of water when flying over their target position.
	\item \emph{Water collection} \: When the water payload is less than 30\% of full capacity, the drone's target is changed to the water reservoir location (top left corner).
	\item \emph{Recharging} \: When the battery charge is low, the drone's target is changed to the location of the docking/charging station (bottom left corner).
	\item \emph{Check} \: After all fires are believed to be suppressed, the swarm will explore the domain searching for new hot spots.
	\item \emph{Landing} \: The nearest (six) drones have priority to fly towards their original docking stations and land. When landed, the collision-avoidance algorithm is deactivated and their positions will not be considered any more for the collision count.
	\item \emph{Hovering} \: The farthest drones will fly towards a waiting area, hovering until there is no more drones with higher priority for landing.
	\item \emph{Landed} \: After every drone has landed, the simulation will continue for another minute to check that all fires are properly extinguished and that no re-ignition occurs.
\end{itemize}

While the first and the last phases are fixed, the others are change dynamically and can coexist. That is to say, some drones may be suppressing fires while others are flying to replenish water or recharge batteries.

Various numerical experiments have been performed with different combinations of maximum cruise speed ($v$) and sampling frequency of the signal field ($f$). For each experiment, the total number of collisions ($C$) and the duration of the mission ($T$) were recorded. A few representative cases are provided in Table~\ref{tab-results}. It is important to note that every drone survives any collision in these numerical experiments. Although this may sound unrealistic, it is done puposely because it is useful for statistical purposes: the rate of collisions would decrease due to a shrinking swarm rather than due to a successful collision-avoidance algorithm.

% Uncomment the following figure if want 2x3 subfigures
%\begin{figure}[h] \label{fig-phases}
%	\begin{subfigure}[b]{0.32\linewidth}\centering
%		\includegraphics[width=0.99\linewidth]{fig-experiment-0s.png} \caption{Initial condition}
%	\end{subfigure}
%	\begin{subfigure}[b]{0.32\linewidth}\centering
%		\includegraphics[width=0.99\linewidth]{fig-experiment-12s.png} \caption{Water collection}
%	\end{subfigure}
%	\begin{subfigure}[b]{0.32\linewidth}\centering
%		\includegraphics[width=0.99\linewidth]{fig-experiment-76s.png} \caption{Refuelling}
%	\end{subfigure} %\hfill
%	\begin{subfigure}[b]{0.32\linewidth}\centering
%		\includegraphics[width=0.99\linewidth]{fig-experiment-127s.png} \caption{Check search}
%	\end{subfigure} \hspace{0.009\linewidth}
%	\begin{subfigure}[b]{0.32\linewidth}\centering
%		\includegraphics[width=0.99\linewidth]{fig-experiment-175s.png} \caption{Hovering and landing}
%	\end{subfigure} %\hspace{0.009\linewidth}
%	\begin{subfigure}[b]{0.32\linewidth}\centering
%		\includegraphics[width=0.99\linewidth]{fig-experiment-199s.png} \caption{Landed} 
%	\end{subfigure} %\hfill
%	\caption{Firefighting phases for a swarm of 100 drones.}
%\end{figure}

% Uncomment the following figure if want 3x2 subfigures

\begin{table}
	\centering
	\renewcommand{\arraystretch}{1.5}
	\newcolumntype{C}[1]{>{\centering\arraybackslash}p{#1}}
	\caption{Experimental results for various combinations of cruise speed ($v$) and sampling frequency ($f$), where $C$ is the total collision count and $T$ is the duration of the simulation.}\label{tab-results}
	\begin{tabular}{|C{0.5cm}|C{1.2cm}|C{0.8cm}|C{0.8cm}|C{0.8cm}|C{1.2cm}|C{1.2cm}|}
		\hline
		ID & $v$ [m$\cdot $s$^{-1}$] &  $f$ [Hz]& $C$ [-] & $T$ [s] & $f \cdot v^{-1}$ [m$^{-1}$]& $C \cdot T^{-1}$ [s$^{-1}$]\\
		\hline
		1 & 5  & 30 & 0 & 551 & 6.0 & 0.00\\
		2 & 10 & 30 & 7 & 264 & 3.0 & 0.03\\
		3 & 15 & 30 & 244 & 182 & 2.0 & 1.34\\
		4 & 20 & 30 & 1019 & 169 & 1.5 & 6.03\\
		5 & 20 & 40 & 243 & 167 & 2.0 & 1.46\\
		6 & 20 & 50 & 36 & 159 & 2.5 & 0.12\\
		7 & 30 & 80 & 1 & 127 & 2.7 & 0.01\\
		\hline
	\end{tabular}
	\renewcommand{\arraystretch}{1}
\end{table}

Experiments 1 to 4 were carried out with the same sampling frequency (30~Hz), showing that increasing cruise speed (5, 10 and 15~m/s) exponentially deteriorates the performance of the proposed collision-avoidance algorithm. In turn, experiments 4 to 6 were carried out with the same cruise speed (20~m/s), showing that increasing sampling frequency (30, 40 and~50 Hz) dramatically decreases the collision count. Whilst the available computational resources allowed us to carry out experiments up to $f = 80~\text{Hz}$, it is reasonable to expect that $f = 100~\text{Hz}$ would bring the collision rate to zero in this particular case. Further experiments will be performed in the near future to support this claim.

By studying the values of the frequency-to-velocity and collisions-to-time ratios (last two columns in Table \ref{tab-results}), we identify a hyperbolic trend: the collisions-to-time ratio tends to infinity for decreasing frequency-to-velocity ratios, whilst the collisions-to-time ratio tends to zero for increasing frequency-to-velocity ratios. It is hypothesised that there is a function that describes the relation between these two coefficients. This is in line with the results from experiments 3 and 5 carried out with the same frequency-to-velocity ratio ($f \cdot v^{-1}$) and returning similar collisions-to-time ratios ($C \cdot T^{-1}$). Further experiments will be carried out in order to better investigate this hypothesis.

\section{Conclusions}

In this paper, we proposed and developed a reactive multi-agent decentralised stigmergy-based algorithm for reciprocal collision-avoidance. Its potential was demonstrated by implementing it on a swarm of drones self-organised to fight the propagation of simulated wildfires. Arguably, the algorithm could also be used for blind flight --i.e. when the drones have no means of knowing the positions of others. The effects of two main parameters were investigated, namely the cruise speed and the sampling frequency, reaching the expected conclusion that decreasing the former and/or increasing the latter results in the reduction of the collision rate. Furthermore, by analysing the frequency-to-speed and collisions-to-time ratios, we hypothesised that there might be a trend that should be further investigated. Furthermore, a convenient by-product of the proposed collision-avoidance algorithm has been observed: it helps maintain diversity in the swarm, thus enhancing exploration.

Purposely, no uncertainty in the environment, no stochasticity in the algorithms, and no flight dynamics were considered in order to focus on the impact of the stigmergy-based collision-avoidance algorithm on the collective behaviour and performance of the swarm. Future work will introduce these aspects as well as extend the developments to three-dimensional space.

\bibliographystyle{splncs04}
\bibliography{Grasso_Innocente_Multi-agent_collision_avoidance}

\end{document}